%% file: manuscript_miska_balzani.tex
\documentclass[parskip = half,twoside,footinclude = false]{scrartcl}
\pdfoutput=1 

\usepackage{ifthen} 

\newcommand{\figpath}{.}
\newcommand{\tikzpath}{.} 

\newboolean{boldeqnitalic}
\setboolean{boldeqnitalic}{true}
\usepackage{defdb1} 
\usepackage{rublkm_paper} 
\usepackage{rotating}

\markauthor{Niklas Miska, Daniel Balzani}
\marktitle{Reliability-Based Design Optimization Incorporating Extended Optimal Uncertainty Quantification}

\begin{document}

\thispagestyle{empty}

\input{sec_01_introduction_and_abstract}

\input{sec_02_theory}

\input{sec_03_computational_example}

\input{sec_04_conclusion}

\section*{Acknowledgement}

The authors greatly appreciate financial funding by the German Science Foundation (Deutsche Forschungsgemeinschaft, DFG) as part of the priority program ``Polymorphic uncertainty modeling for the numerical design of structures'' (SPP 1886), project BA2823/12-2.
Furthermore, the authors thank Markus Wagner for valuable discussions and hints regarding the local laser-hardening processes.
Additionally, the authors thank Hendrik Haddenhorst for his assistance during the data preparation of the considered finite element simulations.

\bibliographystyle{references_style}
\bibliography{zotero}

\end{document}

%% file: sec_01_introduction_and_abstract.tex
\begin{center}
{\LARGE Reliability-Based Design Optimization Incorporating Extended Optimal Uncertainty Quantification}

\vspace{5mm}

Niklas Miska$^{1}$, Daniel Balzani$^{1\star}$

\vspace{3mm}

{\small $^1$Chair of Continuum Mechanics, Ruhr-Universit\"at Bochum,\\
Universit\"atsstra{\ss}e~150, 44801~Bochum, Germany}\\[3mm]

\vspace{3mm}

{\small ${}^{\star}$E-mail address of corresponding author: daniel.balzani@ruhr-uni-bochum.de}

\vspace{10mm}

\begin{minipage}{15.0cm}
\textbf{Abstract}\hspace{3mm}
Reliability-based design optimization (RBDO) approaches aim to identify the best design of an engineering problem, whilst the probability of failure (PoF) remains below an acceptable value.
Thus, the incorporation of the sharpest bounds on the PoF under given constraints on the uncertain input quantities strongly strenghtens the significance of RBDO results, since unjustified assumptions on the input quantities are avoided.
In this contribution, the extended Optimal Uncertainty Quantification framework is embedded within an RBDO context in terms of a double loop approach.
By that, the mathematically sharpest bounds on the PoF as well as on the cost function can be computed for all design candidates and compared with acceptable values.
The extended OUQ allows the incorporation of aleatory as well as epistemic uncertainties, where the definition of probability density functions is not necessarily required and just given data on the input can be included.
Specifically, not only bounds on the values themselves, but also bounds on moment constraints can be taken into account.
Thus, inadmissible assumptions on the data can be avoided, while the optimal design of a problem can be identified.
The capability of the resulting framework is firstly shown by means of a benchmark problem under the influence of polymorphic uncertainties.
Afterwards, a realistic engineering problem is analyzed, where the positioning of laser-hardened lines within a steel sheet for a car crash structure are optimized.
\end{minipage}
\end{center}

\medskip{}
\textbf{Keywords:} reliability-based design optimization, optimal uncertainty quantification, epistemic and aleatory uncertainties, laser-hardened car components

\section{Introduction}

Most engineering problems are affected by uncertainties in one way or another, e.g., by uncertain material properties or uncertain loading scenarios.
Hence, the incorporation of these uncertainties in the search for the optimal design of the engineering problem at hand is an important factor in order to obtain a reliable solution.
For this purpose, methods from the field of reliability-based design optimization (RBDO) have been developed, cf. e.g.,~\citet{EneSor:1994:ros},  \citet{FraMau:2003:lro}, \citet{SchJen:2008:cmo}, \citet{BeaJenSchVal:2013:rou}, \citet{ValSch:2010:sar}, \citet{Goet:2017:neu} or \citet{MacCayEdlFreHanMahMesPen:2019:occa}.
Such RBDO-methods aim to identify an optimal design, assessed by means of suitable performance indicators, whilst a pre-defined level of safety is maintained.
Thus, the chosen performance indicator acts as cost function for the optimization procedure, whilst the desired safety level is usually implemented as constraint to the design optimization, i.e.~all design candidates not meeting the desired level are discarded.
Albeit the definition of safety is not necessarily unique or distinct, the probability of failure (PoF) is a common choice for the evaluation of the problem's safety.\\
With these considerations, RBDO-methods in general require the computation of two different quantities, because the chosen performance indicator for the cost function is mostly not used to assess the PoF for the particular problem.
Consequently, under the presence of uncertainties, both quantities will become uncertain and thus, a framework for uncertainty quantification is required, which is efficient on one hand, because it has to be evaluated twice for each design candidate in the optimization, but also robust on the other hand, as the design optimization relies on the quantification of the mathematically sharpest bounds of these two quantities.
Since in this work polymorphic uncertainties are considered, i.e., both aleatory and epistemic uncertainties, both, the performance measure and the PoF, can only be quantified to lie within certain bounds.
In this context, we consider aleatory uncertainties to be those uncertainties, which are described by probability density functions (PDFs) and additional data, e.g. in form of additional measurements which do not reduce the variability.
Epistemic uncertainties on the other hand are all uncertainties, which are induced by a lack of knowledge and could in principle be reduced, if additional data were available.
Epistemic uncertainties may thus be given in terms of e.g., intervals, fuzzy numbers, sequences of lower order moments or imprecise probabilities.\\
The extended Optimal Uncertainty Quantification (OUQ) framework is a reasonable choice for the required uncertainty quantification as it is able to compute the desired sharpest bounds on the probabilistic event of interest, whilst both epistemic and aleatory uncertainties can be considered.
The original OUQ approach~\citet{OwhScoSulMcKOrt:2013:ouq,MckOwhScoSulOrt:2012:tou,MckStrSulFanAiv:2012:baf} was mainly designed for epistemic uncertainties and has already been applied to engineering problem including the assessment of rupture probabilities in atherosclerotic arteries, see~\citet{BalSchOrt:2017:mft}.
The approach has been extended in~\citet{MisBal:2022:mft} to allow for the incorporation of polymorphic, i.e. epistemic and aleatory uncertainties.
The extended OUQ is also applicable for problems involving spatially distributed properties in the sense of a random field, cf., e.g.,~\citet{Vor:2008:ssc} and~\citet{GhaSpa:1991:sfe}, which requires a nested application of the OUQ, cf.~\citet{MisFreBal:2024:nou}.
In contrast to alternative methods for polymorphic uncertainty quantification, such as fuzzy-randomness or imprecise probabilities, cf. e.g.,~\citet{MoeGraBee:2000:fsa,MoeGraBee:2003:sao} or \citet{BeeFerKre:2013:ipi}, the extended OUQ allows to consider not only bounds on epistemic uncertainties, but also bounds on the moments of the uncertain quantity without the necessity of specifying an underlying type of distribution functions, which are parameterized in terms of the specified moments.
By that, uncertified assumptions on unavailable data such as e.g., moments of higher order implied by the type of distribution functions, can be avoided during the uncertainty quantification.
This does not only allow for an uncertainty quantification without potentially unjustified assumptions on the data, it also enables the systematic investigation, which information on uncertain data such as ranges, bounded or precise moments has to be at least available to reach the defined design targets.\\
In this paper, we propose to include the extended OUQ framework in an outer RBDO-context in terms of a double loop algorithm.
The framework is first tested by means of a benchmark problem as posed in~\citet{PapDauDriDudEhrEicEigGotGraGraGruHomKalMosPetStr:2019:aad}.
Therein, the cross-section of a column is optimized, such that the PoF caused by buckling of the column does not exceed a specified threshold.
The results obtained therefrom are compared to results of related approaches considering e.g., fuzzy-probabilities.
Subsequently, the proposed RBDO framework is used to optimize a car front bumper, for which three different production and lifetime scenarios are investigated: the deep-drawing, a spring-back analysis with prior trimming of the sheet metal and a simplified crash scenario.
The degrees of freedom are positions of locally laser-hardened traces, by which the local material properties such as the yield strength of the sheet metal can be significantly improved, cf. e.g.,~\citet{WagJahBeyBal:2016:dos}.
Therein, the material parameters of the unmodified base sheet are considered to be spatially distributed in terms of a random field.
Results show that more meaningful conclusions can be drawn from the RBDO results with reasonable numerical effort such that relevant engineering problems can be analyzed.

%% file: sec_02_theory.tex
\section{Design Optimization under Polymorphic Uncertainties}

In order to integrate the extended Optimal Uncertainty Quantification framework within an outer RBDO-context, first the general concepts of Reliability-Based Design Optimizations and the general ideas of the extended Optimal Uncertainty Quantification framework following \cite{MisBal:2022:mft} are introduced in \secref{sec:rbdo} and \secref{sec:ouq}, respectively.
Afterwards, the combination of both and the required algorithms are discussed.

\subsection{Reliability-Based Design Optimization}\label{sec:rbdo}

As the name indicates, reliability-based design optimization seeks for the optimal design of a problem under at least one reliability constraint.
As discussed in the introduction, the reliability of any problem at hand is usually assessed by means of the probability of failure (PoF).
However, since the problem may not only be affected by aleatory uncertainties, i.e. the uncertainties can be properly expressed with probability density functions (PDF), but also by epistemic uncertainties, i.e. only intervals or fuzzy numbers for certain parameters, the PoF can generally not be computed as a precise scalar value, rather than to lie within an interval only.
Thus, a framework capable of polymorphic uncertainty quantification is necessary to perform an RBDO under the influence of both types of uncertainties, with the same argumentation holding for the optimization target, i.e. the cost function, as well.
Usually, such frameworks include an inner numerical integration for the incorporation of the aleatory uncertainties and an outer optimization to account for the interval-like nature of epistemic uncertainties.
Based on these premises, the general problem of a reliability-based design optimization in the context of this work can formally be written as
\eb \label{eq:rbdo}
\min_{\Btheta } K(\Btheta, \bz) \quad \mbox{s.t.} \enskip C_j(\Btheta, \bz) \le 0,
\ee
wherein $K$ denotes the cost function, $\Btheta$ represents the vector of optimization parameters, i.e. the degrees of freedom of the optimization problem, and $\bz \defeq  [\by, \widehat{\by}]$ is the joint vector of epistemic~$\by$ and aleatory~$\widehat{\by}$ uncertain quantities influencing the engineering problem~$M$.
As mentioned above, the cost function needs to account for the presence of the polymorphic uncertainties and thus, contains an optimization to identify the most unfavorable value of the model due to the epistemic uncertainties and a numerical integration for the aleatory uncertainties, i.e.
\eb
K(\Btheta, \bz) \defeq \max_{\by} \left[ \mathbb{E}_{\widehat{\by}}(M(\Btheta, \bz))\right].
\ee
As it can be seen, in this work, the expectation~$\mathbb{E}$ has been selected to quantify the impact of probabilistic quantities on the model~$M$, which is no necessity of the method.
Depending on the specific application, other probabilistic measures such as e.g. a quantile may be used.
The mentioned numerical integration is hidden in the computation of~$\mathbb{E}$, since the presence of non-linear models and multiple aleatory uncertainties usually do not permit the analytical computation of the expectation~$\mathbb{E}$.

In addition to the cost function, at least one constraint to the optimization has to be considered, which is the reliability constraint.
This constraint is formally given by
\eb
C_1(\Btheta, \bz) \defeq \max_{\by} \left[ \mu(g(\Btheta, \bz)) \right] - \mathbb{P}_{\textrm{adm}},
\ee
wherein $\mathbb{P}_{\textrm{adm}}$ denotes the maximum admissible value of the PoF and $\max_{\bz} \left[ \mu(g(\Btheta, \bz)) \right]$ poses a second optimization problem in order to identify the largest bound on the PoF of the problem due to the presence of uncertainties.
$\mu(g(\Btheta, \bz))$ itself denotes the computation of the probability of the generalized event $g(\Btheta, \bz)$, here denoting the event of failure, which again requires a numerical computation.

This setting of RBDO requires the computation of two extreme bounds on probabilistic quantities.
Therefore, the utilization of the extended Optimal Uncertainty Quantification (OUQ), as described in the following section, is a meaningful choice for the uncertainty quantification.
By that, the mathematically sharpest bounds on the probability of interest can be computed, while no unjustified assumptions on the data describing the uncertainties have to be made.
In particular, the OUQ allows the incorporation of moment constraints of certain quantities without the requirement of selecting a specific distribution type, while other quantities can be described by intervals, imprecise or precise distribution functions.

\subsection{Extended Optimal Uncertainty Quantification}\label{sec:ouq}

Before the extended OUQ framework as published in~\cite{MisBal:2022:mft} can be discussed, first the definitions of both aleatory and epistemic uncertainties and the resulting algorithmic implications are discussed.
For a polymorphic uncertainty quantification, it is necessary to discriminate both types of uncertainties due to the varying amounts of available data for the individual uncertain quantities.
Whilst some uncertain quantities are well represented by PDFs, for others the available data may not be sufficient and only imprecise representations can be utilized.
Thus, for a generic uncertainty quantification problem of $q$~uncertain quantities, the available data for a single uncertain quantity~$y^{(m)}$ (with $m \in 1, \dots, q$) may include its range, i.e., $y^{(m)} \in [y^{(m)}_{\text{lower}}, y^{(m)}_{\text{upper}}]$, bounds on certain moments, i.e. $\IE[(y^{(m)})^{b}] \in [c_{\text{lower}}^{(m,b)}, c_{\text{upper}}^{(m,b)}]$ for an imprecise moment of order~$b$, precisely known moments, i.e. $\IE[(y^{(m)})^{b}]=c^{(m,b)}$ or, in the case of sufficient knowledge, a probability density function $f^{(m)}(\widehat{y}^{(m)})$.
In the scope of this work, those uncertainties, which can be characterized by a precise probability density function, are referred to as aleatory uncertainties, whilst all other uncertainties are grouped as epistemic uncertainties.
Note that similar to \secref{sec:rbdo}, aleatory uncertainties are denoted by a hat for an easier differentiation against epistemic uncertainties, e.g., $\widehat{y}^{(m)}$.
Accordingly, the vector~$\widehat{\by}$ is a vector of all $s$~aleatory uncertainties of the problem, whilst~$\by$ is the vector of the remaining~$(q-s)$ epistemic uncertainties and $\bz=[\by, \widehat{\by}]$ is the vector of all $q$~uncertainties.
Based on the associated diverse amounts of data on different uncertain quantities, the extended OUQ allows the computation of the mathematically sharpest bounds on a probabilistic event of interest, e.g., the PoF, without the premise of assumed imprecise PDFs for epistemic uncertainties, by e.g. utilizing intervals for characteristic parameters of these PDFs.
In general, the computation of these sharpest bounds can be formulated as the optimization problems
\eb \label{eq:optimal_bounds_inf}
\begin{aligned}
  {\cal L} & \defeq \mathop{\text{inf}}\limits_{\mu \in {\cal A}} \mu \left[ g \le 0 \right],\\
  {\cal U} & \defeq \mathop{\text{sup}}\limits_{\mu \in {\cal A}} \mu \left[ g \le 0 \right],
\end{aligned}
\ee
wherein $g \le 0$ denotes the event of failure and $\mu$ is a multivariate probability measure contained in the set ${\cal A}$.
This set is the set of admissible scenarios, i.e, it contains all those probability measures, which agree with the available data on the uncertain quantities.
Although this set restricts the possible choice of a probability measure~$\mu$ due to the data constraints, the optimization problems in~\eqsref{eq:optimal_bounds_inf} are still optimization problems of infinite dimension, since an infinite number of probability measures complying with the data constraints can be constructed.
Thus, the problems in~\eqsref{eq:optimal_bounds_inf} are numerically unsolvable.
In the context of the OUQ, the multivariate probability measure is composed from one dimensional probability measures for each uncertain quantity, i.e., $\mu = \mu^{(1)} \otimes \dotsc \otimes \mu^{(m)} \otimes \dotsc \otimes \mu^{(q)}$, by which also stochastic independence is induced.
For aleatory uncertainties, the necessary probability measure~$\mu^{(m)}$ can be computed rather easily, since the associated probability density is known due to the definition of an aleatory uncertainty.
The probability measures for the epistemic uncertainties, however, are not as straightforward, since, as mentioned above, multiple measures may comply with the available, limited data on the individual quantity.
Therefore, the reduction theorem from~\cite{OwhScoSulMcKOrt:2013:ouq} is applied, which states, that the optimizers of~\eqsref{eq:optimal_bounds_inf} can be found, if the probability measures for these epistemic uncertain quantities are constructed as convex combinations of Dirac masses.
For the single quantity~$y^{(m)}$, a convex combination of Dirac masses is given by
\eb \label{eq:single_dirac_measure}
\mu^{(m)} = \sum_{k=0}^{n^{(m)}} w^{(m)}_k \cdot \delta(y^{(m)} - y^{(m)}_k),
\ee
wherein $n^{(m)}$ denotes the number of terms of the convex combination, $w^{(m)}_k$ is the weight of the $k$-th term and $\delta(y^{(m)} - y^{(m)}_k)$ represents the corresponding Dirac function centered at its support point~$y^{(m)}_k$.
Since \eqsref{eq:single_dirac_measure} is required to be a valid probability measure, the weights need to be normalized, i.e., $\sum_{k=0}^{n^{(m)}} w^{(m)}_k =1.0$, and $w^{(m)}_k \in [0.0, 1.0]$.
The number of necessary terms $n^{(m)}$ for the convex combination depends on the amount of available knowledge for the particular quantity.
If only the range should be included, a single term is sufficient, i.e, $n^{(m)} = 1$; for every additional moment constraint to be considered, a further term has to be added, regardless if this constraint is related to an imprecise or precise moment.\\
Given the approximation of the probability measures for the epistemic uncertainties by means of the convex combination of Dirac masses, the joint multivariate probability measure can be written as
\eb \label{eq:ouq_pof_epis}
\mu \left[ g(\by) \le 0 \right] = \sum_{i=0}^{n^{(1)}} \dotsc \sum_{k=0}^{n^{(m)}} \dotsc \sum_{l=0}^{n^{(q-r)}} w_i^{(1)} \dotsc ~w_k^{(m)} \dotsc ~w_l^{(q-r)} \chi \left( \by_{i \dotsc k \dotsc l} \right),
\ee
wherein $\by_{i \dotsc k \dotsc l} = [y_i^{(1)}, \dots, y_k^{(m)}, \dots, y_l^{(q-r)}]$ is the vector of the support points fo the Dirac masses for the particular combination~${i \dotsc k \dotsc l}$.
This vector is passed to the function~$\chi$, which evaluates the PoF, i.e.~$g(\bz) \le 0$.
This computation incorporates the aleatory uncertainties of the problem.
Hence, in function~$\chi$, a numerical integration of the failure region of the involved PDFs~$f^{(s)}$ is performed, i.e.
\eb \label{eq:num_integration}
\chi = \displaystyle \int_{g(\bz) \le 0} \displaystyle \prod_{s=1}^r f^{(s)}(\widehat{y}^s) \text{d}\widehat{\by}.
\ee
Usually, this integration is carried out using Monte Carlo approaches such as e.g.,~\citet{deAPatBee:2015:als} or \citet{PapStr:2021:cls}.
It may be noted, that also in the extended OUQ, a combination of epistemic and aleatory uncertainties is possible by e.g., specifying a moment constraint on characteristic parameters of a PDF.

With \eqsref{eq:ouq_pof_epis} in conjunction with \eqsref{eq:num_integration}, the optimization problems formulated in \eqsref{eq:optimal_bounds_inf} can be numerically solved.
Due to the application of the reduction theorem, the number of degrees of freedom is reduced from infinity to a finite number.
Therein, the degrees of freedom are the support points and weights of the constructed convex combinations of Dirac masses for the epistemic uncertainties.
Constraints to the optimizations of \eqsref{eq:optimal_bounds_inf} are firstly bound constraints on the support points and weights of the constructed Dirac combinations.
However, in addition to that, non-linear constraints are necessary.
At first, the normalization constraints on the weights for each epistemic quantity need to be ensured, otherwise no valid probability measure can be constructed.
Furthermore, it has to be assured, that the constructed probability measures exhibit moments within the specified range, i.e. any data constraint must not be violated.
For this purpose, the classical and central moments of order~$b$ of a constructed convex combination of Dirac masses can be computed by
\eb
\label{eq:moments}
\begin{array}{ll}
 \IE_{\mu_m} \left[ \left(y^{(m)}\right)^{b} \right] & = \sum_{k=1}^{n^{(m)}} w^{(m)}_k  \left(y^{(m)}_k\right)^{b} \quad \text{and}  \\[3mm]
 \IE_{\mu_m} \left[ \left(y^{(m)}-\IE_{\mu_m}\left[y^{(m)} \right]\right)^{b} \right] & = \sum_{k=1}^{n^{(m)}} w^{(m)}_k \left(y^{(m)}_k - \IE_{\mu_m}\left[y^{(m)} \right]\right)^{b}, \text{respectively.}
\end{array}
\ee
For an increasing number of uncertain quantities and possibly a high number of moment constraints, which should be incorporated in an uncertainty quantification, the number of non-linear constraints increases and the optimization problem becomes difficult to solve.
Therefore, sophisticated solvers for non-convex and global optimization problems such as Differential Evolution have to be employed, cf. e.g.,~\citet{StoPri:1997:dea} as well as \citet{MckStrSulFanAiv:2012:baf,MckOwhScoSulOrt:2012:tou} for a python implementation in the OUQ context and \citet{Pol:2017:lwc} for an extended version with self-tuning hyper-parameters.
However, it has been found in~\cite{MisBal:2022:mft}, that even then the mathematically sharpest bounds are hard to obtain.
As possible way out, the optimization problem can be reformulated in the space of so-called canonical moments, cf.~\citet{DetStu:1997:tto}, \citet{SteGamKelIos:2020:ouq} and~\cite{MisBal:2022:mft}.
After the reformulation, the degrees of freedom are moments in the classical moment space (only if moments in bounds are specified as input data) and canonical moments of higher order.
From these moments, a matching convex combination of Dirac masses can be computed, which is then used as described above to evaluate the probability of interest.
Although this transformation introduces additional computational steps, the overall efficiency of the optimization is significantly increased, since only bound constraints remain and the entire search space becomes admissible.\\
In the context of RDBO, not only the sharpest bounds on the probability of failure are of interest.
For the evaluation of a design candidate also the sharpest bound on the performance measure is important, so that the cost function can be computed as well.
In analogy to the computation of the moments of a single convex combination, the mean of a model response~$M$ can be computed by
\eb
\mathbb{E}(M(\bz)) =\sum_{i=0}^{n^{(1)}} \ldots \sum_{j=0}^{n^{(q-r)}} w^{(1)}_i \ldots w^{(n)}_j\,\chi(M(y^{(1)}_{i}, \ldots, y^{(n)}_{j})).
\ee
By that, an objective function similar to \eqref{eq:ouq_pof_epis} is obtained and can be used to identify the mathematically sharpest lower or upper bound on the mean under the given uncertain quantities.

\subsection{Integration of the Extended OUQ in an RBDO-Context}
\begin{figure}[t]
\begin{center}
  \includegraphics{\tikzpath/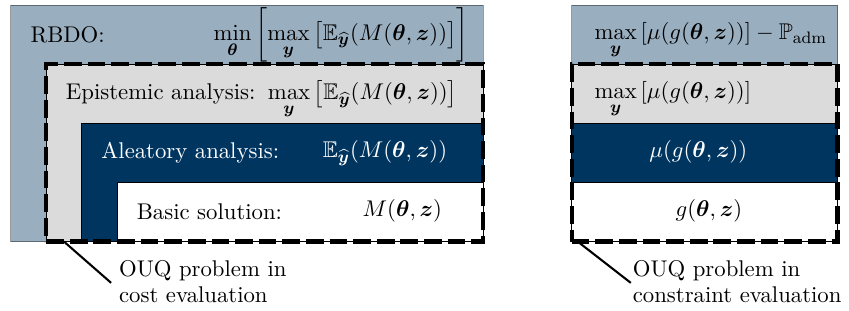}
\end{center}
\caption{Integration of the extended OUQ framework within a RBDO context with the necessary steps for the evaluation of the cost function on the left and for the reliability constraint on the right.}
\label{fig:rbdo}
\end{figure}

The integration of the extended OUQ into an RBDO-framework leads to a nested optimization or a double loop approach, cf.~also the diagram in \figref{fig:rbdo} and the simplified algorithm for the design objective function in Alg.~\ref{alg:rbdo_nested}.
The outer optimization is concerned with the design optimization and the degrees of freedom are the design parameters~$\Btheta$.
The inner optimizations, focus on the uncertainty quantification in terms of the extended OUQ.
However, since not only the upper bound on the expectation of the model response, i.e., $\max_{\by} \left[ \mathbb{E}_{\widehat{\by}}(M(\Btheta, \bz))\right]$, but also the reliability constraint and thus, the upper bound on the PoF have to be computed, two optimizations for a single design candidate are necessary.
Therefore, the efficiency of the implementation of the extended OUQ is a crucial part, as numerous individual uncertainty quantifications are required over the course of a design optimization.\\
The above description may imply a differentiation between design parameters~$\Btheta$ and uncertain quantities~$\bz$.
However, the double-loop approach allows for a combination of both, i.e.~a dependency of an uncertain quantity from a design parameter, e.g.~$y^{(m)}(\theta^{(j)})$.
A setting like this may be for instance meaningful for the determination of an admissible uncertainty for a quantity whilst a desired target with respect to the cost can be met.
Since the uncertainty quantification is performed in the inner optimizations, the dependency $y^{(m)}(\theta^{(j)})$ will be constant for these two optimizations, however, for different design candidates the shape of the uncertainty will change.
In practice, the design candidate could be, for example, an interval midpoint, i.e. $y^{(m)} \in \left[ \theta^{(j)} - w/2, \theta^{(j)} + w/2\right]$ with $w$ being the interval length.
In the context of OUQ, this approach is not limited to the range of uncertain quantities itself, it is also possible to formulate these interval constraints for moments in bounds for epistemic uncertainties.
In general, it could be possible to select from different types of PDFs based on a design parameter~$\theta^{(j)}$.
In this work, however, only the interval midpoint is utilized in the second numerical example.
\begin{algorithm}[t]
\caption{Cost function for RBDO}
\label{alg:rbdo_nested}
\DontPrintSemicolon
\SetInd{0.4em}{0.8em}
\KwData{Design candidate $\Btheta_m$, max. bound on PoF $\mathbb{P}_{\text{adm}}$, uncertainties $\bz$}
\KwResult{cost value $K$}

    update $\bz(\Btheta)$ if dependency exist\;

    compute value for cost $K = \max_{\by} \left[ \mathbb{E}_{\widehat{\by}}(M(\Btheta, \bz)) \right]$\;

    compute PoF $\mathbb{P} = \max_{\by} \left[ \mu(g(\Btheta, \bz)) \right]$\;

    \;

    \If(){$\mathbb{P} > \mathbb{P}_{\text{adm}}$}{
        discard candidate $\Btheta_m$\;
    }

    \;

    \If(){$K < K_{\text{best}}$}{
        $K_{\text{best}} = K$\;
        $\Btheta_{\text{best}} = \Btheta_m$\;
    }
\;
\Return{$K$}
\end{algorithm}

%% file: sec_03_computational_example.tex
\section{Numerical Examples}

This section is devoted to the application of the combined framework to two examples.
The first example is a benchmark problem posed for optimizations under the influence of polymorphic uncertainties, cf.~\cite{PapDauDriDudEhrEicEigGotGraGraGruHomKalMosPetStr:2019:aad}, which has been published in the context of the priority program~1886 (SPP1886) of the German Research Foundation.
The analytic nature of the underlying problem allows the convenient comparison of different frameworks or approaches for the uncertainty quantification, since expensive numerical simulations in terms of e.g., finite elements are avoided.
In the second case, a more complex problem is investigated, wherein the placement of locally laser-hardened zones on exemplary car front bumper is optimized.
This example includes the numerical simulation of three production steps and thus, relies on the construction of surrogate models in order to keep the numerical cost of the RBDO in a feasible range.
For both examples, the design optimization and the uncertainty quantification are implemented using the programming language Julia, cf.~\citet{BezKarShaEde:2012:jfd}.
If not specified otherweise, the optimizations are performed based on the LSHADE44 algorithm, cf.~\citet{Pol:2017:lwc}, which is a variant of the Differential Evolution method~\cite{StoPri:1997:dea}, an evolutionary algorithm able to solve global non-convex optimization problems.
In contrast to the original method, the LSHADE44 algorithm uses different mutation methods simultaneously and self-tunes its hyper-parameters in order to improve the convergence behavior towards the optimal solution.
The examples are computed on nodes containing an Intel Xeon Phi 7210 with 64 cores of up to 1.50 GHz core frequency and 92 GB RAM.

\subsection{Benchmark of SPP1886}
The benchmark problem from~\cite{PapDauDriDudEhrEicEigGotGraGraGruHomKalMosPetStr:2019:aad} for polymorphic uncertainties proposes two challenges.
Whilst the first challenge is concerned with the assessment of the PoF only, which has been analyzed for the extended OUQ in~\cite{MisBal:2022:mft}, the second challenge focuses on the optimization of parameters, which were fixed in the first challenge, such that the PoF does not exceed a specified upper bound.
\begin{figure}[t]
\unitlength1cm
\centering
\includegraphics{\tikzpath/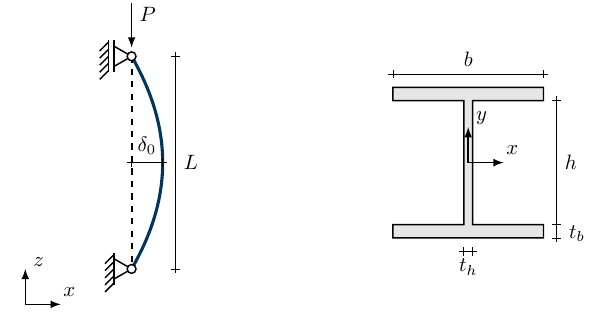}
\caption{Illustration of the benchmark buckling problem as proposed in~\cite{PapDauDriDudEhrEicEigGotGraGraGruHomKalMosPetStr:2019:aad}, taken from~\cite{MisBal:2022:mft}.
\label{fig:benchmark_illust}}
\end{figure}
The structure itself is an H-beam made of steel acting as a column and loaded by a compression force~$P$ on the top, cf.~the illustration of the column and its cross-section in \figref{fig:benchmark_illust}.
Due to its load and cross-section, the failure mode is buckling around the y-axis.
The challenge of the benchmark is the optimization of the cross-section, such that the area of the cross-section and by that, the required amount of material, is minimized, whilst a PoF of $\mathbb{P}_{\textrm{adm}}=1.3 \cdot 10^{-6}$ is not exceeded.
Although the depicted cross-section has multiple characterizing dimensions, only the width~$b$ and height~$h$ are optimized.
Additionally, the ratio of these two dimensions is fixed to $b/h=1$, i.e., only one design parameter remains for the optimization leading to $\Btheta = [b]$.
Based thereon, the geometric properties of the cross-section are computed as
\eb
A=2bt_{b}+ht_{h},~W=\frac{ht_{h}^3}{6b}+\frac{b^2t_{b}}{3},~I=\frac{ht_{h}^3}{12}+\frac{b^3t_{b}}{6},
\ee
with $t_{h}=10~\text{mm}$ and $t_{b}=15~\text{mm}$.
The length of the column is fixed to $L=7.5~\textrm{m}$.
Since the area of the cross-section~$A$ is to be minimized, a special case of the design optimization will be performed where~$A$ is considered to depend only on the design parameter, but not on an uncertain quantity.
Hence, $A = A(\Btheta)$ is a deterministic quantity and an uncertainty quantification is only necessary for the reliability constraint. 
Therefore, the following uncertainties are incorporated:
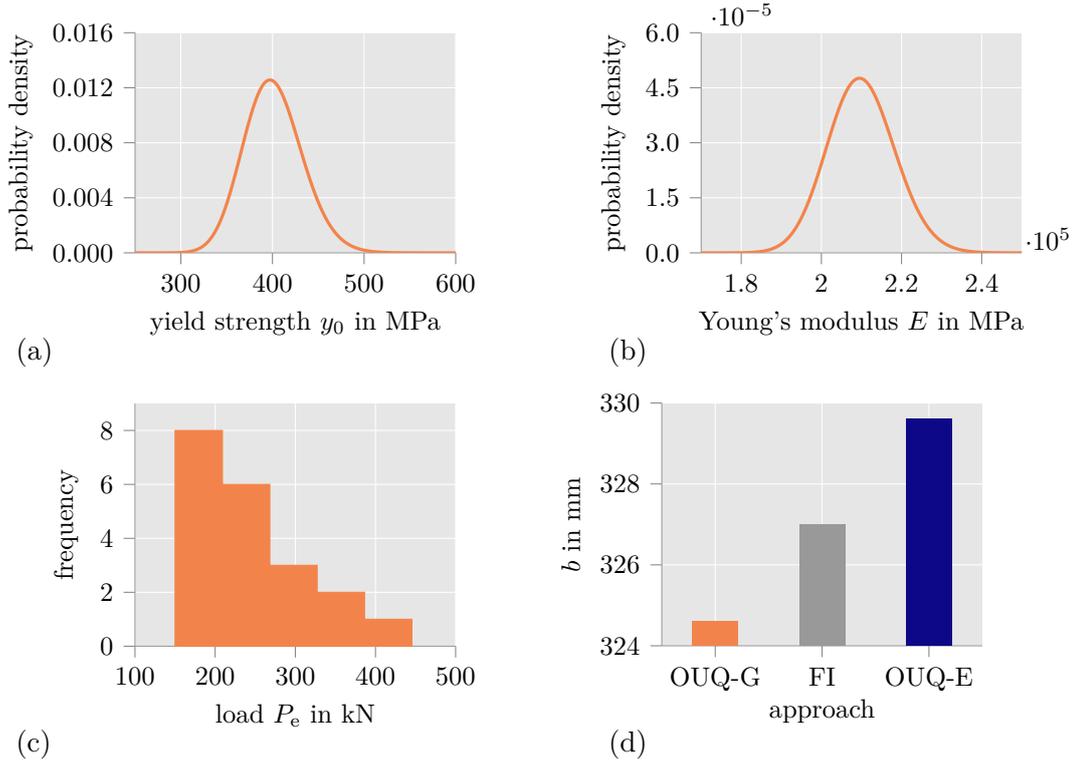
\begin{figure}[t]
\unitlength1cm
\begin{picture}(15.5,9.8)
\put( 0.0, 0.0){
\put( 0.7, 5.4){\includegraphics{\tikzpath/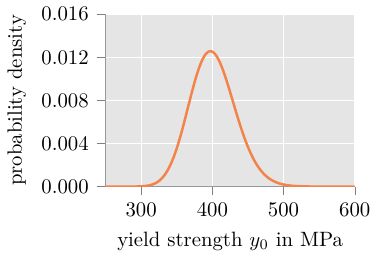}}
\put( 0.92, 5.2){(a)}
\put( 8.5, 5.4){\includegraphics{\tikzpath/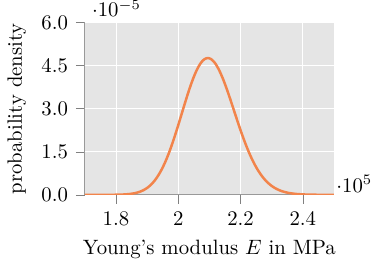}}
\put( 8.72, 5.2){(b)}
\put(1.3, 0.2){\includegraphics{\tikzpath/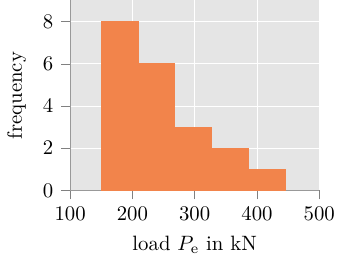}}
\put(0.92, 0.0){(c)}
\put( 7.97, 0.25){\includegraphics{\tikzpath/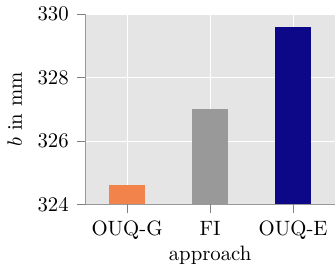}}
\put( 8.72, 0.0){(d)}
}
\end{picture}
\caption{Probability density function for (a) yield strength~$y_0$ and (b) Young's modulus~$E$ and (c) the histogram of the environmental load~$P_e$, taken from~\cite{MisBal:2022:mft}.
In (d), the computed flange width~$b$ is plotted for the two investigated OUQ cases and the fuzzy-interval approach (FI) from~\cite{PapDauDriDudEhrEicEigGotGraGraGruHomKalMosPetStr:2019:aad}. Herein, OUQ-G denotes the imprecise Gumbel distribution function and OUQ-E the approach using only moments in bounds up to order three.
    \label{fig:bench_uncertain}}
\end{figure}
\vspace{-0.2cm}
\begin{itemize}
 \item \textbf{permanent load $P_{\textrm{p}}$}: interval quantity $P_{\textrm{p}} \in \left[100~\textrm{kN}, 200~\textrm{kN} \right]$,

 \item \textbf{environmental load $P_{\textrm{e}}$}: data is given as histogram in the benchmark, two different cases are investigated:
    \begin{enumerate}
        \item Gumbel distribution with parameters in bounds and mean constraint
        \item Moments in bounds up to third order
    \end{enumerate}

 \item \textbf{initial deflection $\delta_0$}: interval quantity $\delta_0 \in \left[0\,\textrm{cm} , 6\,\textrm{cm} \right]$

 \item \textbf{yield strength $y_0$}: log-normal probability density function with a mean of $400\,\textrm{MPa}$ and a standard deviation of $32\,\textrm{MPa}$, as depicted in \figref[a]{fig:bench_uncertain}

 \item \textbf{Young's modulus $E$}: log-normal probability density function, having a mean of $210 000\,\textrm{MPa}$ and a standard deviation of $8400\,\textrm{MPa}$, depicted in \figref[b]{fig:bench_uncertain}
\end{itemize}
With these uncertainties, the event of failure, i.e. buckling due to Euler mode~2, can be computed by means of the limit-state function
\eb
g\left(P_{\textrm{p}}, P_{\textrm{e}}, \delta_0, y_0, E \right) = 1 - \left( \frac{P_{\textrm{p}}+P_{\textrm{e}}}{y_0 A} + \frac{(P_{\textrm{p}}+P_{\textrm{e}}) \delta_0}{y_0 W_{\textrm{y}}} \cdot \frac{P_{\textrm{b}}(E)}{P_{\textrm{b}}(E)-P_{\textrm{p}}-P_{\textrm{e}}} \right).
\ee
If this function yields a negative value, the column will fail due to its loading.
As the posed design optimization problem is linear in its parameters, a gradient descent optimization is used for the outer optimization.
The inner optimization, however, is still non-linear as described for the extended OUQ above, the LSHADE44 optimizer following~\cite{Pol:2017:lwc} and Combination Line Sampling for the Monte Carlo integrations, cf.~\citet{PapStr:2021:cls}, are used.
Since both investigated variants for the environmental load $P_{\textrm{e}}$ require similar numbers of degrees of freedom, in both cases a population size of $n_{\text{pop}}=50$ and a convergence criterion after $100$ iterations for the optimizer as well as $50$ sampling lines for the integration are utilized.
Note that the following calculations were repeated 8 times in order to check if equal optimizers were found to conclude a global optimum.
Indeed, the same results up to the precision given in the provided values were obtained for all 8 calculations.

\subsubsection{Gumbel distribution with imprecise parameters (OUQ-G)}

In the first case a Gumbel distribution with parameters in bounds and with a mean constraints in bounds is incorporated for the environmental load.
Precisely, $a_{\textrm{e}} \in [188.0, 236.0]~\textrm{kN}$ with $\IE[a_{\textrm{e}}]  \in [201.4, 222.6]~\textrm{kN}$ and $b_{\textrm{e}} \in [37.0, 74.0]~\textrm{kN}$ with $\IE[b_{\textrm{e}}]  \in [49.685, 54.915]~\textrm{kN}$ are the intervals for the two characterizing parameters $a_{\textrm{e}}$ and $b_{\textrm{e}}$.
With that, the result of the design optimization is $b=h=324.6\text{mm}$ with an associated area $A=12984\text{mm}^2$.
The averaged computing time over 8 computations is $32$~hours and $56$~minutes.
If the resulting dimension is compared to the result from the fuzzy-interval approach from~\cite{PapDauDriDudEhrEicEigGotGraGraGruHomKalMosPetStr:2019:aad}, here a smaller value has been found, cf.~\figref[c]{fig:bench_uncertain}.
Since all incorporated uncertainties are the same except for the environmental load, the additional information considered here, the mean constraint on the parameters of the Gumbel distribution, is the essential factor.
The mean constraint allows just a little bit less variation of the environmental load, whilst the fuzzy-interval approach accounts for worse cases with only intervals for the parameters and thus, requires a larger cross-section.

\subsubsection{Moment Constraints up to order 3 (OUQ-E)}

For the second case, only moment information deduced from the histogram is incorporated for the environmental load.
In addition to the range $P_{\textrm{e}} \in [100.0, 500.0]~\textrm{kN}$, the following three moments in bounds are used: $P_{\textrm{e}} \in [0.0, 1000.0]~\textrm{kN}$ with $\IE[P_{\textrm{e}}]  \in [209.4, 279.0]~\textrm{kN}$, $\IE[(P_{\textrm{e}}-\IE[P_{\textrm{e}}])^2]  \in [2251.91, 9007.66]~\textrm{kN}^2$ and $\IE[(P_{\textrm{e}}-\IE[P_{\textrm{e}}])^3]  \in [121775, 974204]~\textrm{kN}^3$.
This moment information represents less information on the environmental load than an assumption of a probability density function with imprecise parameters, and thus, the computed solution in this case yields an even larger cross-section than for the fuzzy-interval approach.
In particular, $b=h=329.6\text{mm}$ with $A=13184\text{mm}^2$ is computed, with an average computing time of $28$~hours and $44$~minutes.
This result does not only agree with the different levels of knowledge for this particular quantity, it also agrees with the findings of~\cite{MisBal:2022:mft}.
There, for this second case for fixed geometric parameters a higher PoF was found, whilst the first case led to a smaller PoF than for the fuzzy-interval approach.

\subsection{Optimized Locally Laser-Hardened Traces in Car Front Bumper}

The second investigated example focuses on the design optimization of a car front bumper, for which the optimal placement of locally laser-hardened traces is sought.
The objective therein is the maximization of the dissipated energy of this front bumper in a crash scenario in order to protect the passenger cell in the best possible way.
Local laser-hardening is a technique which allows the targeted improvement of material properties due to a phase change of the sheet metal by a controlled melting and a subsequent controlled cool-down of the steel, cf. e.g.~\citet{WagJahBeyBal:2016:dos}.
For example, the yield stress of a steel base material can be improved by a factor of up to 3.
By that, the sheet metal may be enhanced at specific locations, such that, e.g., a forming process succeeds with a higher probability compared to an unmodified sheet metal.
However, this process has some limitations, as the traces cannot be placed to close to another, i.e. not the entire sheet can be covered.
This is due to the melting and cooling, a second line placed too close to the first line would inevitable alter the properties of the first line again.
Also, the laser-hardening requires a certain processing time, which may become way to high in the case of too many traces.
In these cases, a base sheet made of a higher class of steel may be favorable.
Furthermore, the placed traces are limited in their width, which is mainly caused by the dimensions of the focusing lenses and the required energy of the laser.
Here, the trace width is considered to be limited to $1.5~\text{mm}$ and the sheet can also not be thicker than $1.5~\text{mm}$.
Furthermore, the distance between two traces has to be greater than 1.5 times the trace width.
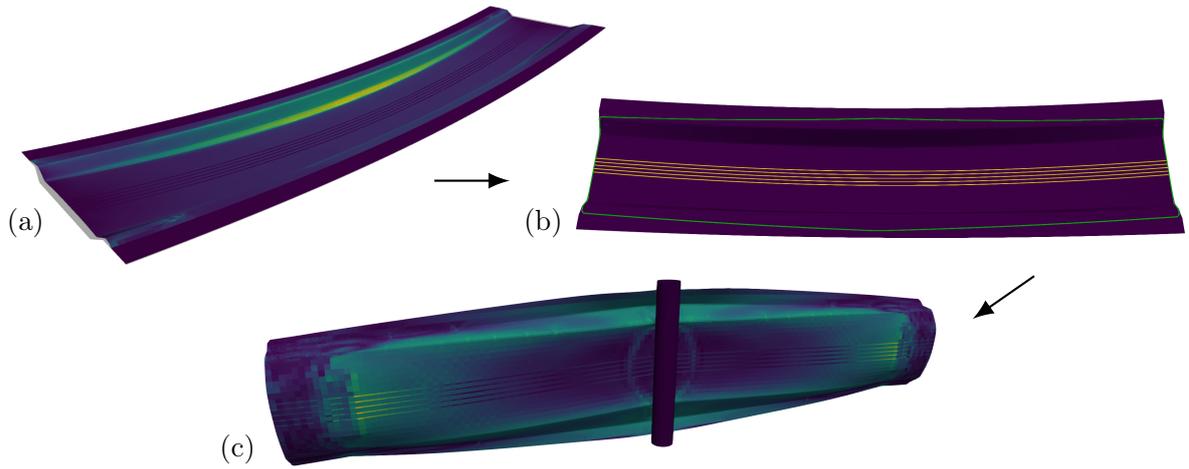
\begin{figure}[t]
\unitlength1cm
\begin{picture}(15.5,5.5)
\put( 0.0, 0.0){
\put( 0.0, 2.5){\includegraphics[width=8.0cm]{\figpath/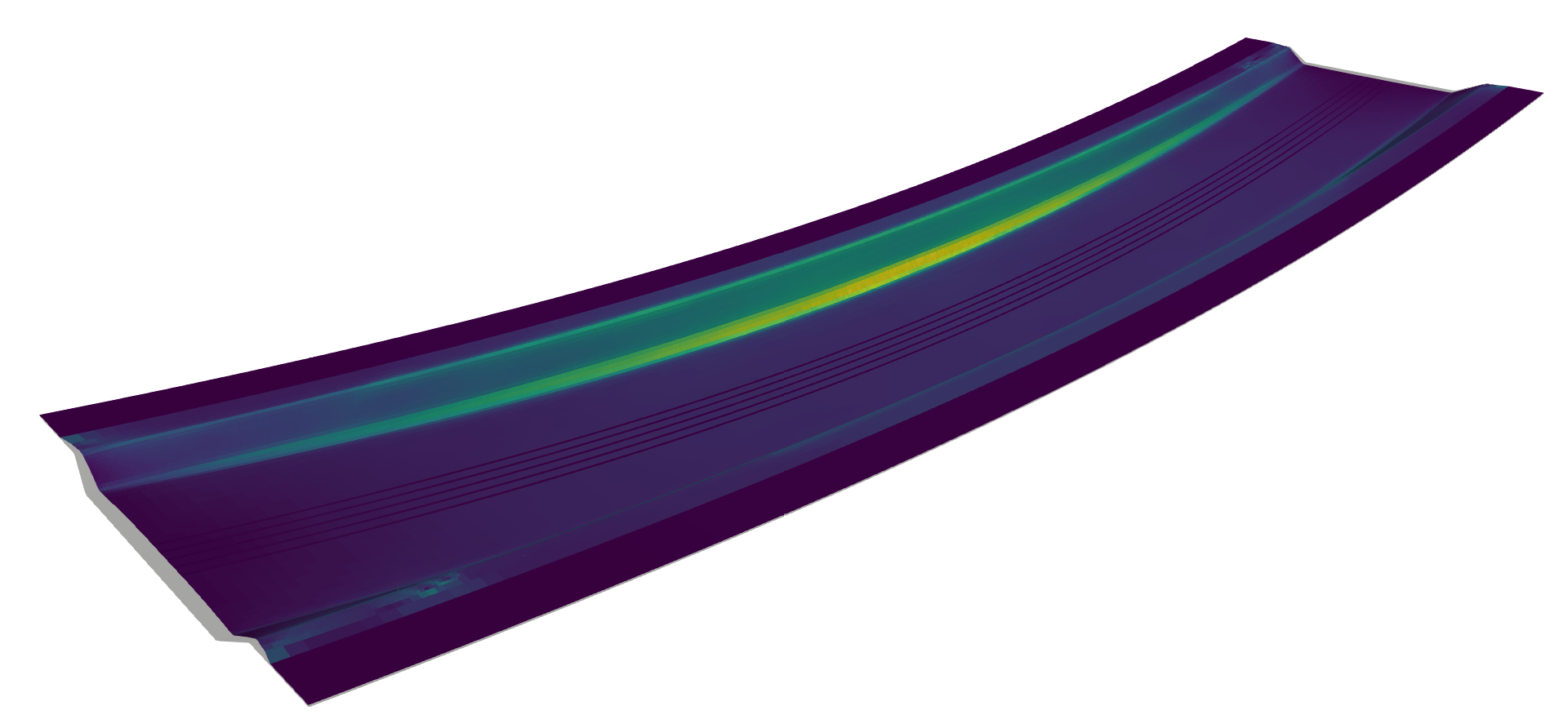}}
\put( 5.1, 3.4){\scalebox{2.0}{$\fancyarrow$}}
\put( 0.0, 3.0){(a)}
\put( 7.5, 2.9){\includegraphics[width=8.0cm]{\figpath/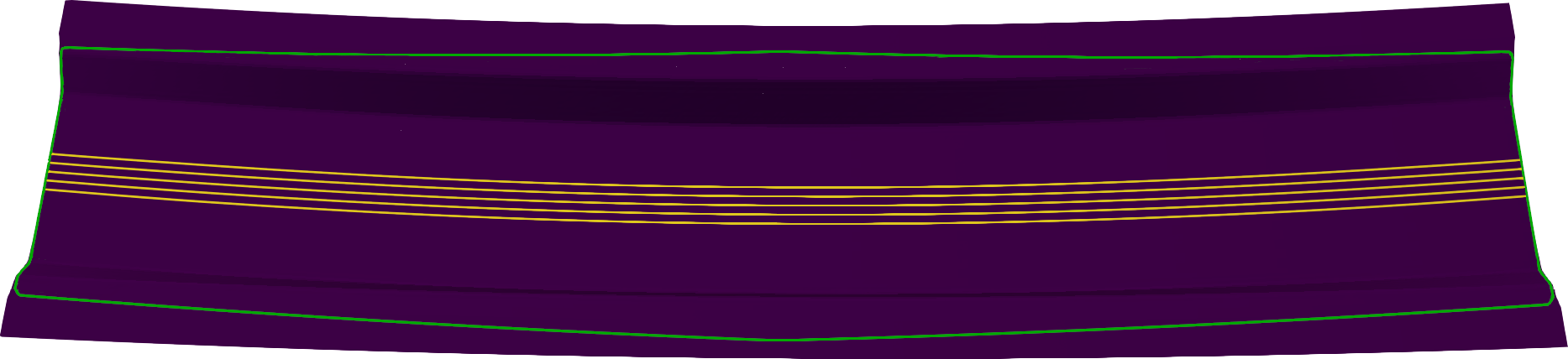}}
\put(13.8, 2.9){\begin{rotate}{-145}\scalebox{2.0}{$\fancyarrow$}\end{rotate}}
\put( 6.8, 3.0){(b)}
\put( 3.3,-0.7){\includegraphics[width=9.0cm]{\figpath/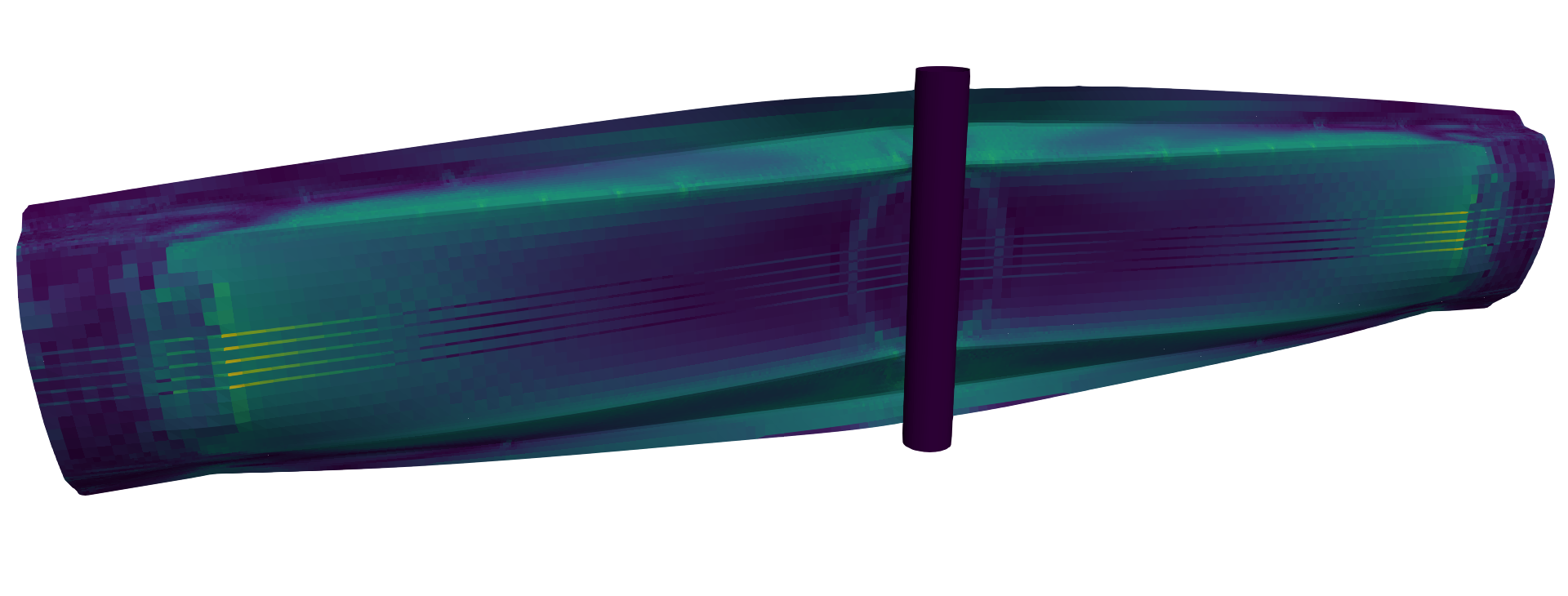}}
\put( 2.8, 0.0){(c)}
}
\end{picture}
\caption{The three investigated stages of simulation: (a)~deep-drawing of the sheet metal, (b)~trimming and spring back and (c)~simple frontal crash. \label{fig:simulation}}
\end{figure}

In our investigation, three different stages are simulated in terms of finite elements: first, the deep drawing process of a sheet metal into the shape of the front bumper is simulated, see \figref[a]{fig:simulation}, to include the eigenstress distributions resulting from the sheet-metal forming process.
Subsequently, the second stage is used for trimming off of excess material and a spring-back analysis, cf. \figref[b]{fig:simulation} resulting in the final configuration of the car component, before in the third step a simple front crash scenario is simulated, \figref[c]{fig:simulation}.
The simulations in terms of finite elements are performed using~LS-Dyna R8.1.0~\cite{lst:2015:lsd}, also on the Xeon Phi nodes specified before.
However, as already a single evaluation of the function~$\chi$, which is used in the extended OUQ to calculate the probability of interest for a deterministic combination of points, requires at least three numerical simulations, an uncertainty quantification or even a design optimization based on actual finite element simulations becomes far to costly from a numerical perspective.
Therefore, surrogate models are employed, which yield approximately the same result as the finite element simulations, but which are magnitudes cheaper to evaluate within the optimization.
In this work, feed-forward neural networks are employed, which are trained as regression models on a set of pre-computed samples.
Since the selection of an appropriate topology is a challenge by itself, the number of layers, the number of neurons per layer and also the activation functions within these neurons have been optimized (trained with hyperband tuner from Tensorflow-Keras-Software).\\
The goal of the design optimization is to maximize the dissipated energy~$D$ during a crash in order to protect the passenger cell of the car in an optimal way.
For this purpose, three different failure modes are considered, each associated to one of the three simulation stages.
Firstly, failure of the sheet metal forming is determined by the Cockroft-Latham criterion, i.e. $ W_{\textrm{C}} = \int_0^{\overline{\varepsilon}} \mathop{\text{max}}(\sigma_1, 0) \text{d}\overline{\varepsilon} \ge W_{\textrm{C}}^{\textrm{ult}}$ with $\overline{\varepsilon}$ being the equivalent plastic strain, $\sigma_1$ the first principial stress and $W_{\textrm{C}}^{\textrm{ult}}$ the ultimate admissible value, cf. e.g.~\citet{BjoLarNil:2013:fhs} or~\citet{TarHopLanClaHilLadEri:2008:slp}.
Secondly, during the spring-back, an upper threshold for the node displacement is specified, by which the stability of the shape is assessed.
Finally, also the deflection of the car front bumper during the crash scenario is limited to include protection of the engine and cooler behind the bumper in a real car.
In order to keep the required work for this example in a feasible range, the example is limited to only straight horizontal traces.
However, in a realistic application, different orientations of different, more complex, trace shapes could be considered.
Nevertheless, the present example is a meaningful, more complex case to demonstrate the capabilities of an RBDO with the integrated extended OUQ.
With the restrictions in mind, the design degrees of freedom are the positions of the laser-hardened traces.
For this purpose, the metal sheet is assigned 38 discrete locations, in which these traces can be positioned, cf. \figref[a]{fig:laser_traces}.
These locations consider the aforementioned geometric constraints such as width and minimum distance between the traces.
Furthermore, exactly five traces are positioned, cf. e.g., \figref[b]{fig:laser_traces} for five traces centered in the middle of the sheet, while \figref[c]{fig:laser_traces} denotes a more distributed positioning of the traces.
\begin{figure}[b]
\unitlength1cm
\begin{picture}(15.5,3.5)
\put( 0.0, 0.0){
\put( 0.0, 0.0){\includegraphics{\tikzpath/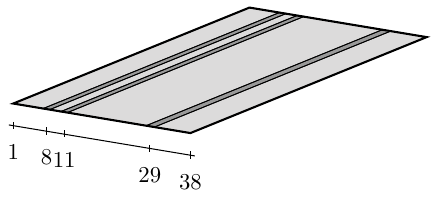}}
\put( 0.0, 0.0){(a)}
\put( 6.0, 0.5){\includegraphics[width=5.0cm]{\figpath/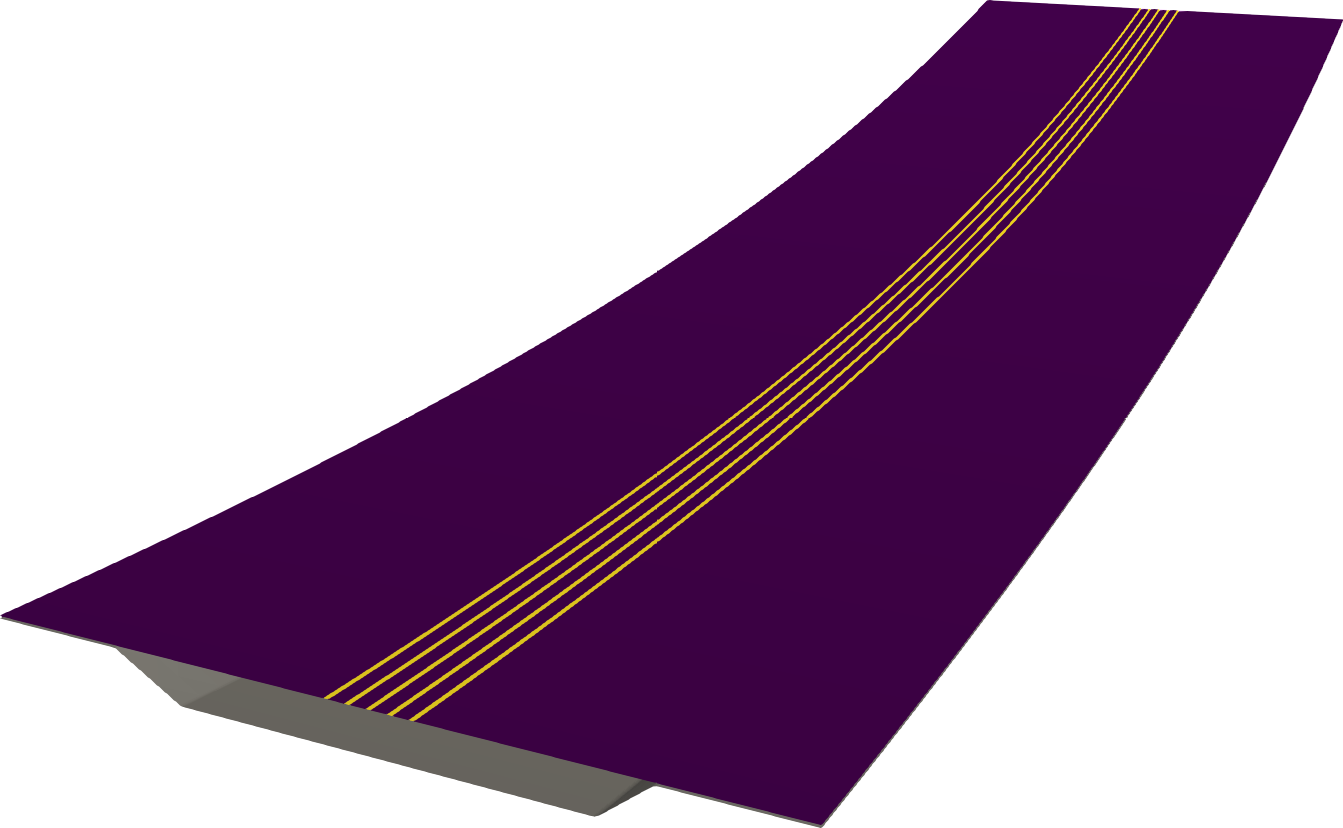}}
\put( 6.0, 0.0){(b)}
\put(10.5, 0.5){\includegraphics[width=5.0cm]{\figpath/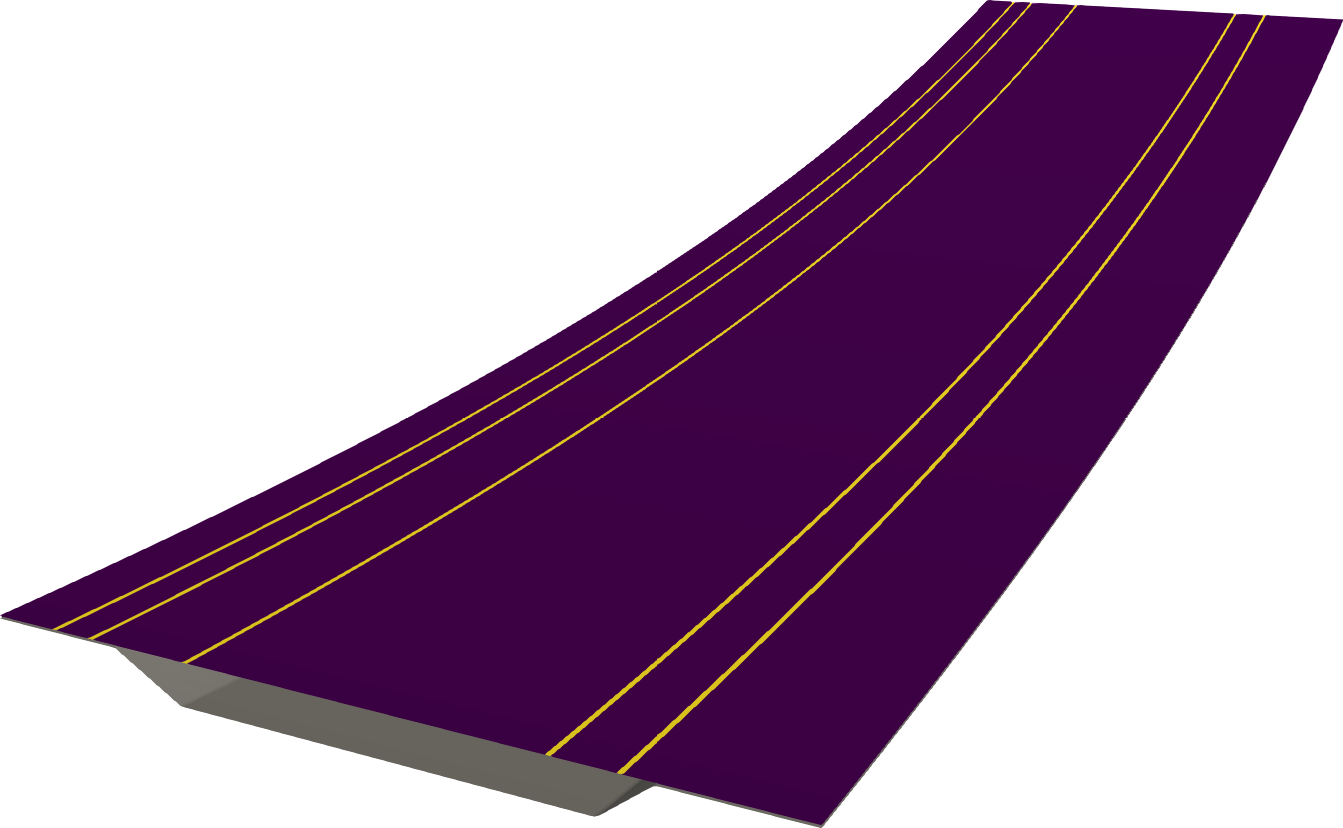}}
\put(10.5, 0.0){(c)}
}
\end{picture}
\caption{(a) Schematic illustration of three exemplary (instead of the five used in the example) laser-hardened traces on the sheet metal with the positions in terms of the range 1 to 38. (b)~and (c)~are two different, possible laser trace positions considering five laser traces. \label{fig:laser_traces}}
\end{figure}
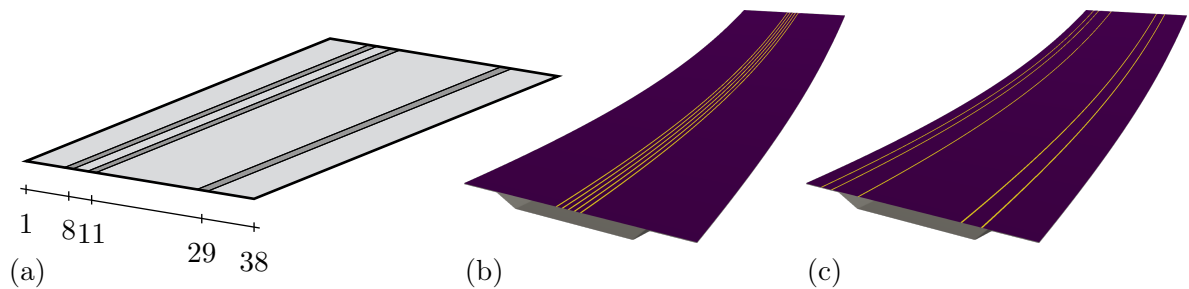
This choice of parametrization of the design optimization, i.e., $\Btheta=[l_1,l_2,l_3,l_4,l_5]$ with $l_i$ being the trace index, leads to an integer optimization.
However, although not being the optimal numerical environment for this task, the design optimization problem is again solved with the LSHADE44 optimizer as described before.
To this end, the parameters are rounded to the closest integer.
Additionally, during the optimization it has to be ensured, that the locations are unique, i.e. a duplicate of a trace index is not admissible due to physical reasons.

\subsubsection{Problem Setting, Uncertain Quantities and Surrogate Models}
\begin{figure}[t]
\unitlength1cm
\begin{picture}(15.5,4.3)
\put( 0.0, 0.0){
\put(-0.2, 0.2){\includegraphics{\tikzpath/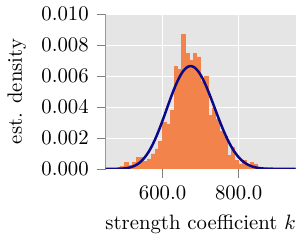}}
\put( 0.0, 0.0){(a)}

\put( 5.25, 0.2){\includegraphics{\tikzpath/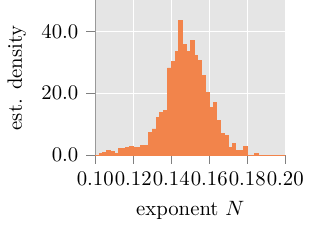}}
\put( 5.35, 0.0){(b)}

\put(10.5, 0.2){\includegraphics{\tikzpath/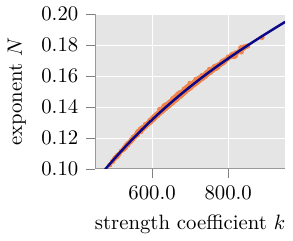}}
\put(10.6, 0.0){(c)}

}
\end{picture}
\caption{Histograms of (a) the strength coefficient~$k$ and (b) the exponential hardening coefficient~$N$ for the used hardening law. Additionally, the beta probability density function for parameter~$k$ is plotted. (c) Correlation of parameters~$k$ and~$N$ with fitted third-order polynomial, cf. \eqsref{eq:correlation_poly}. The data is taken from~\cite{MisFreBal:2024:nou}. \label{fig:material_histogram}}
\end{figure}
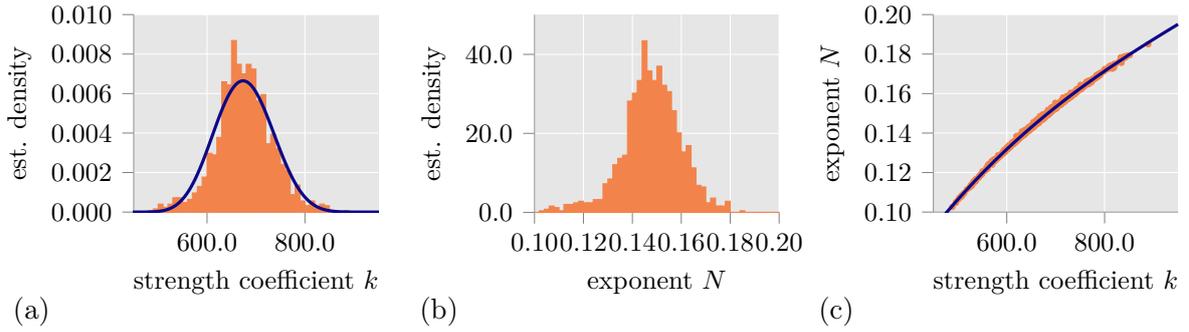

As already mentioned above, in this example the dissipated energy during a simple crash scenario should be optimized.
Thus, the sequence of maximization and minimization as presented in \eqref{eq:rbdo} are inversed:
\eb
\max_{\Btheta } \min_{\by} \left[ \mathbb{E}_{\widehat{\by}}(D(\Btheta, \bz))\right],
\ee
as we want to maximize the worst case, i.e., the minimum expected dissipated energy~$D$, that can be computed due to the line positions~$\Btheta$ and the uncertain quantities~$\bz$.
In this example, the following quantities are considered uncertain:
\begin{itemize}
    \item Material parameters $k$ and $N$ related to the hardening behavior: as identified in~\citet{MisBal:2019:qou} using the automized microstructure simulation approach~\cite{FanMisBal:2020:asv}, the macroscopic hardening behavior of a DP-600 steel depends on the variability of their microstructure's morphology.
    The data is adapted for the Swift-Hardening law within LS-Dyna, \citet{Swi:1952:pip}, which takes the form  $\Bsigma_y = k \left( \Bvarepsilon + \Bvarepsilon_p  \right)^N$.
    Therein, $\Bsigma_y$ denotes the yield stresses, $\Bvarepsilon$ the elastic logarithmic strains and $\Bvarepsilon_p$ the effective logarithmic plastic strains. 
    The parameters $k$ and $N$ influence the shape of the yield curve and are identified based on the data of~\cite{MisBal:2019:qou}, cf. also~\citet{MisFreBal:2024:nou}.
    For both parameters, a histogram is shown in \figref{fig:material_histogram}.
    A correlation analysis reveals a strong correlation between both parameters, for efficiency reasons only the strength coefficient~$k$ is included within the uncertainty analysis, whilst the exponential hardening coefficient is derived from the following polynomial:
    \begin{align}\label{eq:correlation_poly}
        N(k)=&\phantom{+}1.661789\cdot 10^{-10}\,\widehat{k}^3 - 5.0435951923582\cdot 10^{-7}\,\widehat{k}^2 \nonumber  \\
        &+ 6.5798512505658687\cdot 10^{-4}\,\widehat{k} -0.11730087709283931741,
    \end{align}
    in which $\widehat{k}=k \frac{1}{\textrm{MPa}}$ is the unit-less nominal value of the strength coefficient~$k$.
    The correlation and the polynomial are shown in \figref[c]{fig:material_histogram}.
    Additionally, in \figref[a]{fig:material_histogram}, a fitted beta probability density function with $k \in \left[ 400.0, 1000.0 \right] \textrm{MPa}$ and shape parameter $q_1=11.6161$ and $q_2=13.6657$ is plotted.

    \item Material parameter $k_{\text{trace}}$: due to the laser-hardening of the sheet metal, the hardening of the steel changes and thus, the parameter of the considered hardening law.
    Here, it is considered as interval parameter with $k_{\text{trace}} \in \left[ 1200.0, 1800.0 \right] \textrm{MPa}.$

    \item The friction coefficient $f_{\text{c}}$ between sheet metal and tools: in accordance to~\citet{FigRamOliMen:2011:esf}, the friction coefficient is an uncertain quantity, however, only little data is available based on experiments.
    Therefore, the friction coefficient is here modeled as interval quantity with $f_{\text{c}} \in \left[0.1, 0.15 \right]$ and an additional mean constraint $\E[f_{\text{c}}] \in \left[0.105, 0.115 \right]$.

    \item ultimate value $W_{\textrm{C}}^{\textrm{ult}}$ for the Cockroft-Latham criterion: similar to the friction coefficient, only a few experiments are performed for the identification of the ultimate values for the Cockroft-Latham values, cf. e.g.~\cite{TarHopLanClaHilLadEri:2008:slp} or \citet{BjoLarNil:2013:fhs}.
    In~\cite{TarHopLanClaHilLadEri:2008:slp}, values for a DP-800 steel are given, which are here slightly reduced for the DP-600 steel.
    Thus, the ultimate value is modeled as interval quantity: $W_{\textrm{C}}^{\textrm{ult}} \in \left[ 450.0, 550.0 \right] \textrm{MPa}.$
\end{itemize}

For the incorporation of the material parameter~$k$, two different approaches are investigated.
Firstly, the material parameter is considered to be spatially distributed and is hence, modeled by a random field, cf.~\citet{Vor:2008:ssc}, \citet{GhaSpa:1991:sfe} and also~\cite{MisFreBal:2024:nou}.
Since no spatial correlation information is available, every finite element is assigned a random value of~$k$ drawn from its beta distribution, such that in total over the entire sheet a histogram following the beta distribution is obtained.
In the second variant, the strength coefficient~$k$ is assumed to be uniform over the entire sheet, as the fluctuation of the material parameter potentially occurs on a smaller scale than an element size considered here.
Therefore, in that variant, $k$ is an interval quantity with $k \in \left[ 617.35, 734.0 \right] \textrm{MPa}$, which is the range of the standard deviation around the mean of the fitted beta distribution.
Whilst the failure due to the simulation is estimated on the basis of an uncertain quantity, the ultimate values for failure due to spring-back and deformation during the crash are purely deterministic, as these are design specifications.
Here, a maximum value of 30 mm is used for nodal displacement after spring-back and 125 mm of admissible deflection after the crash.
The admissible upper bound on the probability of failure is set to $\mathbb{P}_{\text{adm}}=0.1\%$.\\
Based on the described problem setting, finite element simulations are carried out in order to construct neural networks as surrogate models, cf. also~\citet{Had:2022:mde}.
For this purpose, the range of every parameter serving as input is subdivided in equal parts, such that a set of training data is generated, which spans the entire input space.
Since two variants for the inclusion of the strength coefficient~$k$ are investigated, for each of which three finite element simulations are computed, in total six neural networks are constructed.\\
In the case of including the random field, the inputs for the neural networks are the line positions~$\Btheta$, the friction coefficient~$f_{\text{c}}$ and the strength coefficient~$k_{\text{trace}}$ within the laser-hardened traces.
For every combination of these input parameters, 30 realizations of the random field of the strength coefficient~$k$ are constructed and computed, from which statistics on the quantities of interest are deduced.
In total, 5400 individual settings with 3 finite element simulations each are performed.
The 30 deduced maximum values for the Cockroft-Latham criterion $\max W_{\textrm{C}}$, which is later compared to $W_{\textrm{C}}^{\textrm{ult}}$, and the the maximum node displacements after the spring-back, which result for a fixed input parameter combination from the 30 performed random field computations, are ordered in increasing order and are assigned a normalized frequency of occurrence.
By that, an estimated cumulative distribution function is created, cf. also~\cite{MisFreBal:2024:nou} for details, which can be used as estimate for the percentage of the random fields, which exhibit critical values for the maximum Cockroft-Latham criterion and the maximum node displacement.
Thus, an additional parameter for the neural nets is $W_{\textrm{C}}^{\textrm{ult}}$ or the maximum admissible value for the node displacement, respectively, and the return value is the evaluated value of the constructed empirical cumulative distrbution function.
The neural net for the maximum dissipated energy is constructed a little bit differently.
There, the return value is the mean of the dissipated energy of all 30 random fields, i.e., $\mathbb{E}_{\widehat{\by}}(D(\Btheta, \bz))$.
In the case of uniform properties, also 5400 individual settings are computed, however, here no random fields had to be constructed and incorporated in the neural network.
Hence, the neural networks are straightforward regression fits, with the inputs of the line positions~$\Btheta$, the friction coefficient~$f_{\text{c}}$ and the strength coefficient~$k_{\text{trace}}$ within the laser-hardened traces.
Depending on the computation of interest, the return value of the networks is either the maximum value for the Cockroft-Latham criterion~$\max W_{\textrm{C}}$, the maximum node deformation, or the dissipated energy~$D$ during the crash.\\
For now, the design parameters and the uncertain quantities are clearly separated and no dependency exists.
Therefore, an additional case is investigated to answer the question: can increased knowledge on the strength coefficient~$k_{\text{trace}}$ within the traces improve the optimization result?
The increased knowledge can for example be interpreted as an improved production process, in which the process of the laser-hardening is better monitored to achieve more consistent results.
Here, a mean constraint on~$k_{\text{trace}}$ is introduced to replicate this improved knowledge, additionally, the midpoint of the mean in bounds should be optimized for an improved design.
In doing so, an interval width of 200 MPa is allowed, i.e., $\mathbb{E}[k_{\text{trace}}] \in [\theta_k - 100\textrm{MPa}, \theta_k + 100\textrm{MPa}]$, at which $\theta_k$ denotes the interval midpoint.
With that, the vector of design parameters becomes $\Btheta=[l_1,l_2,l_3,l_4,l_5, \theta_k]$.
The constructed surrogate models do not need to be adapted, since the interaction is not applied to a random field parameter, and can be used as before.

\subsubsection{Optimization 1: Material Parameter as Random Field}
\begin{figure}[t]
    \unitlength1cm
    \begin{picture}(15.5, 3.2)
    \put( 0.0, 0.0){
\put(0.0, 0.5){\includegraphics[width=0.45\textwidth]{\figpath/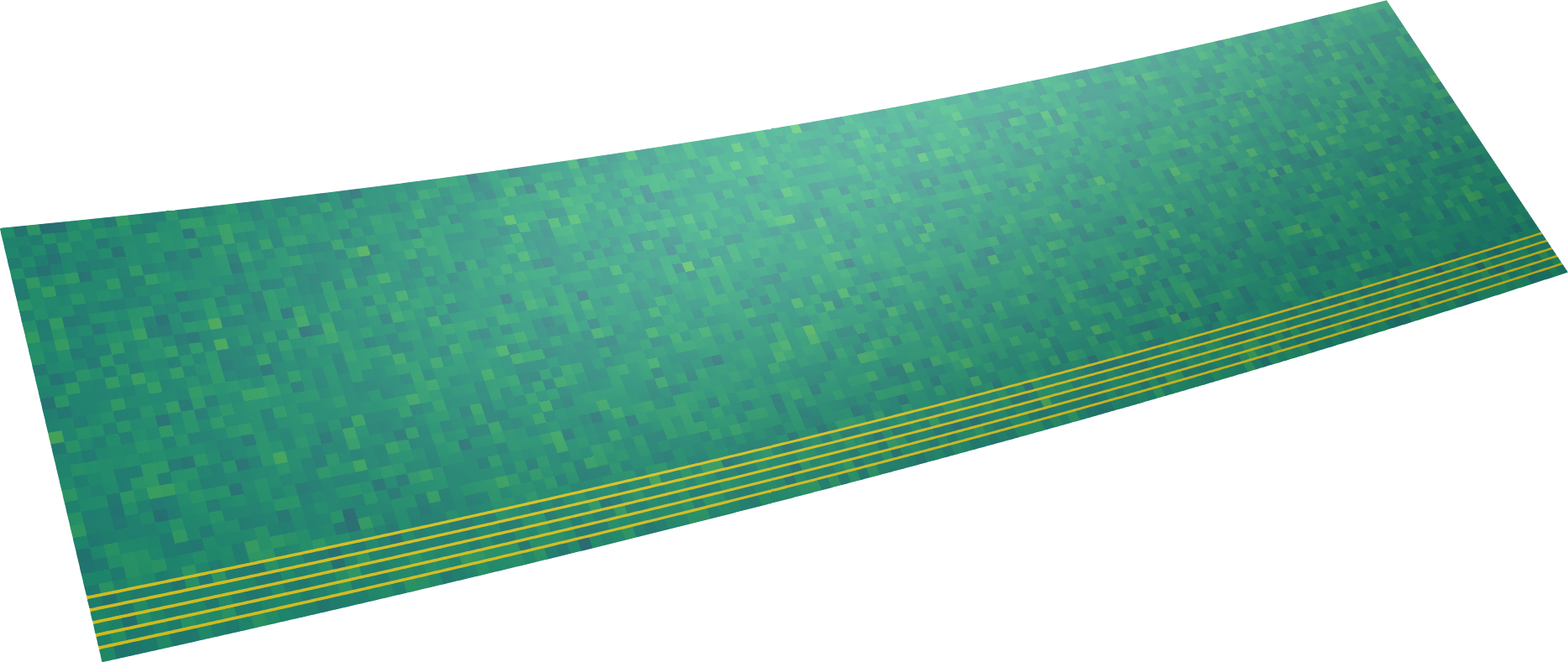}}
\put(0.0, 0.0){(a)}

\put( 8.0, 0.5){\includegraphics[width=0.48\textwidth]{\figpath/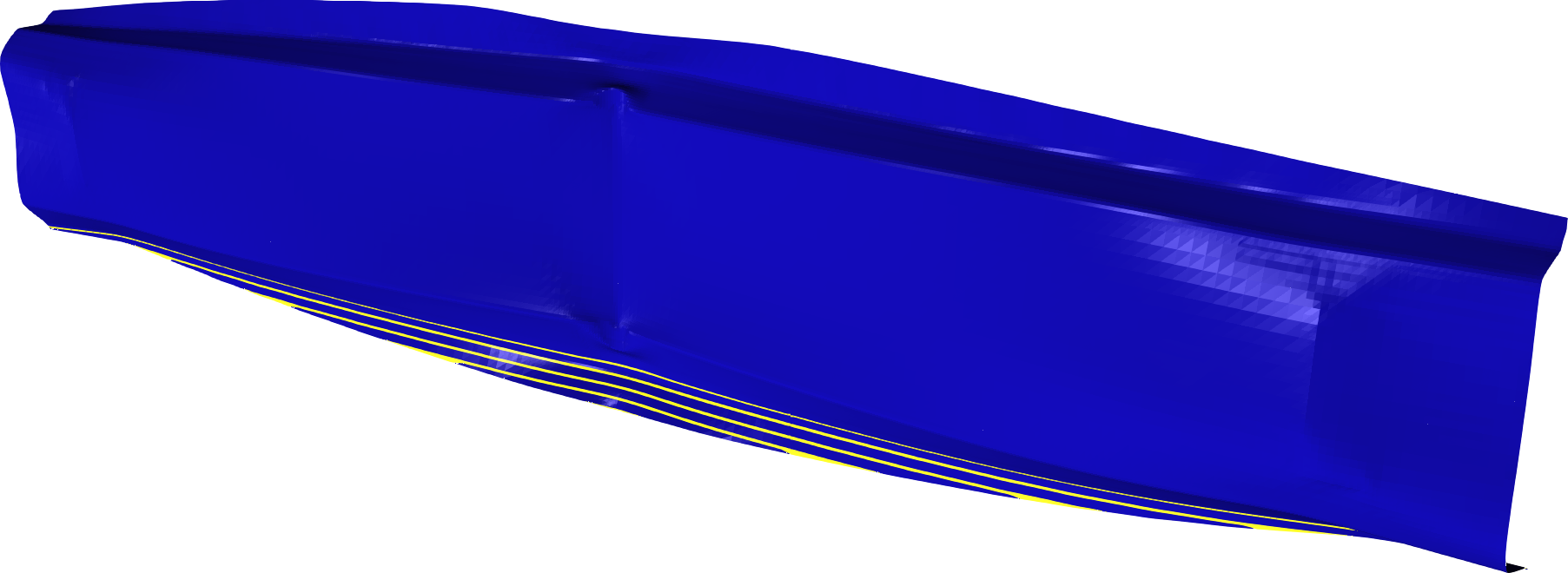}}
\put( 8.0, 0.0){(b)}
}
\end{picture}
\caption{(a) Optimization result for the bumper with random-field and laser-hardened traces mostly at the bottom of the sheet, (b) the deformation of the same sheet after the crash simulation.
\label{fig:forming_res_rf}}
\end{figure}
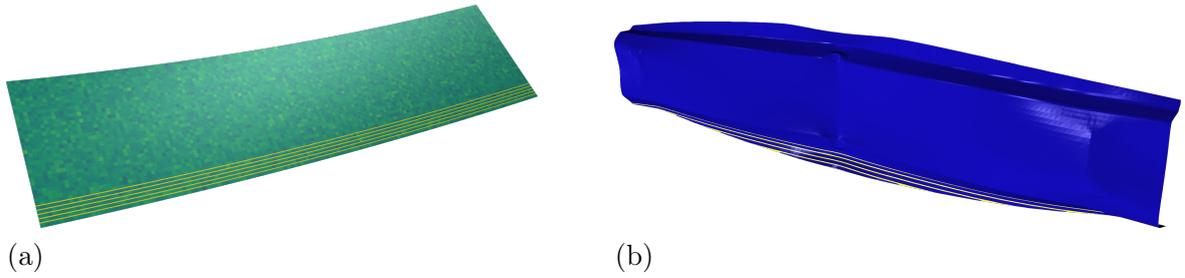

With the described problem setting, the design optimization can be carried out.
Since both, the design optimization and the uncertainty quantifications, are non-convex global optimization problems, the LSHADE44 optimizer is applied.
For the inner uncertainty quantifications, a population size of 250 with a convergence criterion of 250 iterations is used, while the design optimization uses a population of 128 with 50 iterations for convergence.
Exploiting the nature of the evaluations, the calculation of different design candidates within a generation can easily be parallelized.
Here, 64 design candidates were evaluated in parallel on the 64 cores of the used node.
For the first case without interaction between design parameters and uncertaint quantities, the following combination of traces is found optimal:
\eb
\Btheta_{\text{opt}}=[26, 29, 30, 31, 32],
\ee
which leads to a dissipated energy of $D=490,864$ Joule.
In \figref{fig:forming_res_rf}, a training sample for the random field with similar trace positions is depicted together with its deformed state after the crash.
As can be seen, the high indices denote positions of all traces close to the bottom of the sheet metal.\\
With the addition of the mean constraint~$\mathbb{E}[k_{\text{trace}}]$, the optimizer settings remain unchanged, which leads to the following optimal parameter vector
\eb
\Btheta_{\text{opt}}=[26, 30, 31, 32, 33, 1700.0],
\ee
which in turn results in a dissipated energy of $D=491,863$ Joule.
Compared to the result without the constraint, only a $0.2\%$ improvement of the dissipated energy can be realized.
Most likely, this small gain can not justify the required effort in the production to ensure the required additional monitoring.
From the result, it can also be seen, that the highest possible midpoint for the interval of the mean constraint was chosen.

\subsubsection{Optimization 2: Uniform Material Parameter in Bounds}
As mentioned previously, the variability of the incorporated material parameter may happen on a smaller length scale as the element size considered here.
Thus, it could be more realistic to consider only the mean of the obtained beta distribution, here a range of one standard deviation around the mean is used, i.e., $k \in \left[ 617.35, 734.0 \right] \textrm{MPa}$.
Since every other problem setting remains the same, the same optimizer settings as before are used.
With that, the optimal positions are found as
\eb
\Btheta_{\text{opt}}=[2, 3, 23, 24, 28],
\ee
with a dissipated energy of $D=485,035$ Joule.
In contrast to before, not all traces are positioned at the bottom, now two groups are formed.
One group at the top of the sheet, and the second group closer to the middle of the sheet, cf. also \figref[a]{fig:forming_res_uni} for a similar training candidate.
However, in total the dissipated energy is less than for the case with the random field.
Also here, the additional mean constraint~$\mathbb{E}[k_{\text{trace}}]$ is also investigated, which again results in a different position vector:
\eb
\Btheta_{\text{opt}}=[2, 3, 23, 24, 28, 1700.0],
\ee
with $D=487,536$ Joule.
As before, only a small improvement of $0.2\%$ in the dissipated energy can be observed.
Also, the midpoint was again the highest possible value, that could be used.
\begin{figure}[t]
    \unitlength1cm
    \begin{picture}(15.5, 3.8)
    \put( 0.0, 0.0){
\put(0.0, 0.5){\includegraphics[width=0.45\textwidth]{\figpath/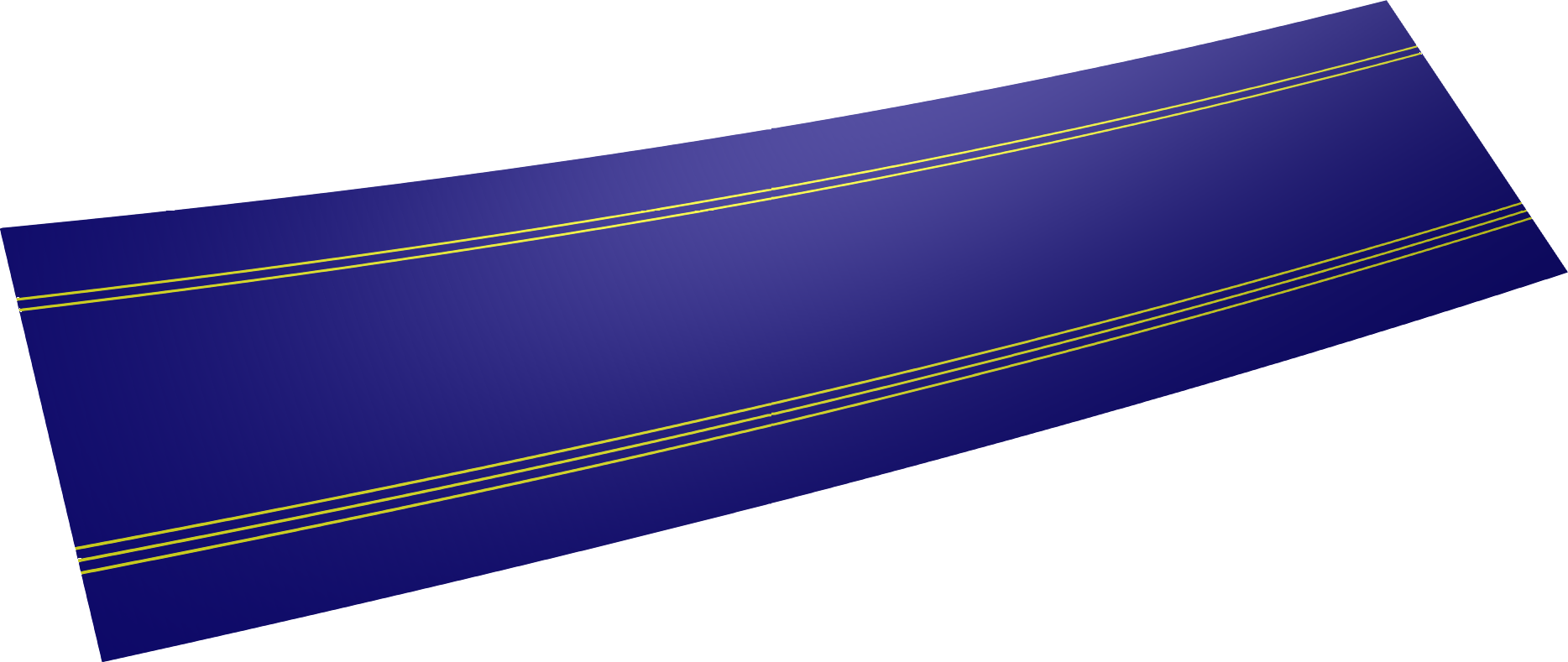}}
\put(0.0, 0.0){(a)}

\put( 8.5, 0.5){\includegraphics[width=0.45\textwidth]{\figpath/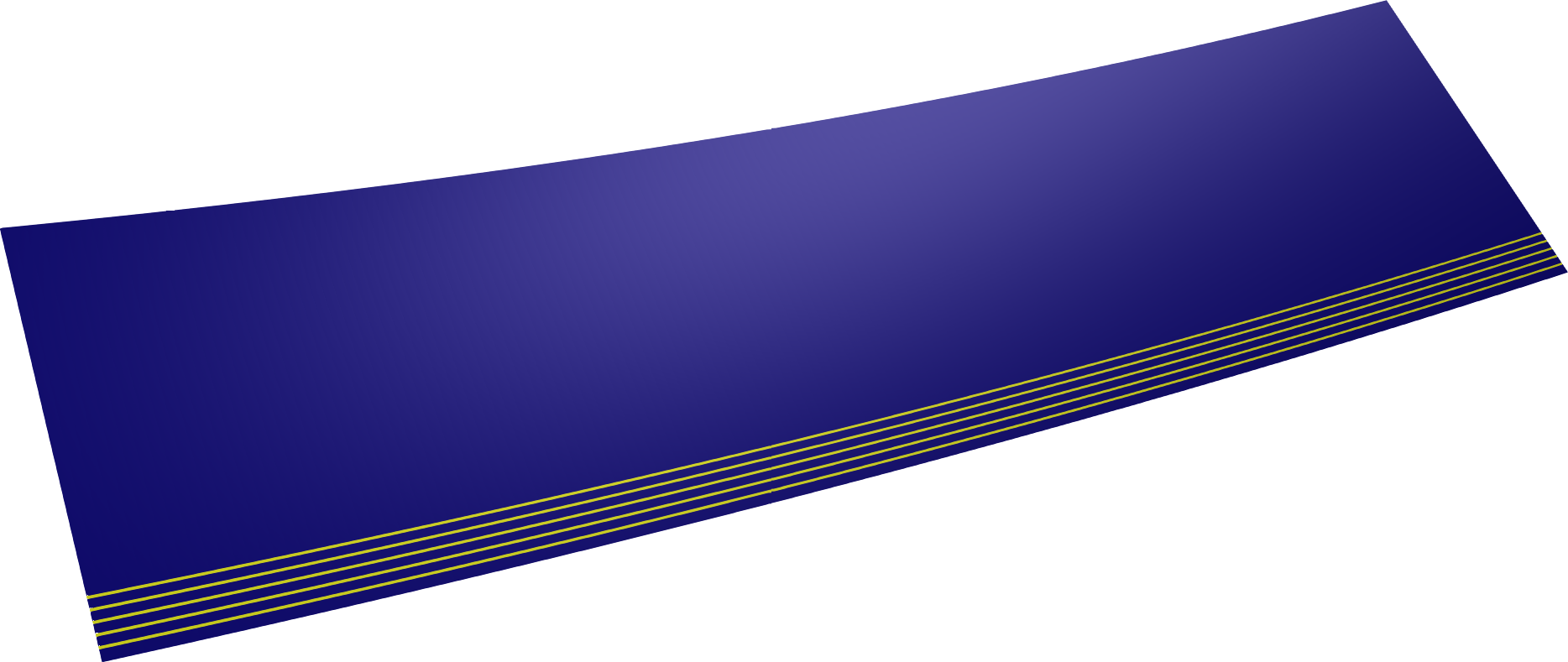}}
\put( 8.5, 0.0){(b)}
}
\end{picture}
\caption{(a) Illustration of the traces for the uniform case, in contrast to (b) the trace locations resulting from the case with random field.
\label{fig:forming_res_uni}}
\end{figure}
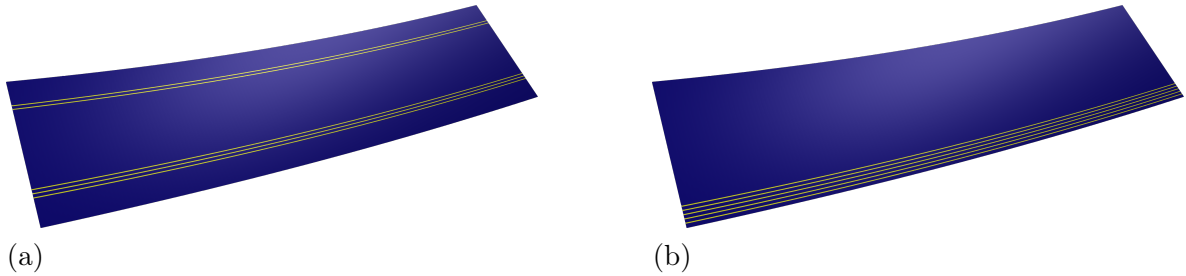
For the presented examples, the investigated dependency between an optimization parameter and an uncertain quantity did actually not result in an improved design.
Nevertheless, the functionality of the proposed RBDO approach based on the extended OUQ could be demonstrated for a problem including a challenging simulation setup.
Thereby, the applicability to practical and computationally demanding engineering tasks has been shown.

%% file: sec_04_conclusion.tex
\section{Conclusion}
A novel approach RBDO problems incorporating the extended OUQ framework in a double-loop-based strategy has been presented in this work.
By that, polymorphic, i.e. both, aleatory and epistemic, uncertainties can be analyzed in the reliability constraint as well as in the objective function of the design optimization.
Therein, the unique capabilities of the extended OUQ are utilized, i.e. it is possible to include moment information on epistemic uncertainties without the necessity of the specification of an underlying probability density function, which would induce also higher moment constraints.
Thus, more robust results can be expected from the design optimization, since only available data is incorporated without any unjustified assumptions.\\
The constructed framework has been tested by means of two numerical examples.
The first example is a benchmark for polymorphic uncertainty quantification, by which the obtained results could be compared to alternative approaches based on fuzzy numbers or intervals for epistemic uncertainties.
Two scenarios with different levels of data on an epistemic uncertainty were investigated.
The obtained results turned out to be plausible in comparison to the results of competitive approaches.
Hence, it can be concluded that the integration of the extended OUQ in an RBDO-context is a meaningful concept.
The second, more complex example is a multi-step simulation of the production process and crash behavior of a simplified car front bumper, in which the placement of locally laser-hardened traces within the sheet metal has been optimized in order to maximize the dissipated energy during the crash.
The challenge herein has been the associated numerical effort, which could be alleviated by employing neural networks as surrogate models for the regression of the model response.
In addition to the first example, a dependency between an optimization variable and an uncertain quantity has been investigated, since the midpoint of an interval describing an uncertainty was intended to be optimized.
Although the latter cases did not result in an improvement of the design objective, the technical integration did work as expected and it was shown that such dependencies can be incorporated within design optimization.
Nevertheless, the examples also outlined, that the design optimization is tightly connected with substantial numerical costs, especially due to the considered double-loop approach.
Hence, a particular focus for further development of this method should be put on increasing efficiency.